\documentclass[11pt]{article}
\usepackage[a4paper,margin=1in]{geometry}
\usepackage[T1]{fontenc}
\usepackage[utf8]{inputenc}
\usepackage{lmodern}
\usepackage{amsmath,amssymb,amsthm,mathtools}
\usepackage{graphicx}
\usepackage{booktabs}
\usepackage{array}
\usepackage{enumitem}
\usepackage{forest}
\usepackage{tikz}
\usetikzlibrary{arrows.meta,positioning,calc}
\usepackage{natbib}
\usepackage[colorlinks=true,citecolor=blue,linkcolor=blue,urlcolor=blue]{hyperref}

\newtheorem{theorem}{Theorem}
\newtheorem{proposition}{Proposition}
\newtheorem{corollary}{Corollary}

\newtheorem{definition}{Definition}
\newtheorem{remark}{Remark}

\DeclareMathOperator{\ETL}{ETL}
\DeclareMathOperator{\ETLCont}{ETLCont}
\newcommand{\MR}{\mathrm{MR}}
\newcommand{\CR}{\mathrm{CR}}
\newcommand{\Gauss}{\mathrm{Gauss}}
\newcommand{\Clay}{\mathrm{Clay}}
\newcommand{\R}{\mathbb{R}}
\newcommand{\1}{\mathbf{1}}
\newcommand{\dd}{\,\mathrm{d}}
\newcommand{\ThetaH}{\Theta}

\title{Explicit Finite-Sum Tail Risk Measures for Hierarchical Market--Credit Copula Aggregation}

\author{Luisa Tibiletti\\
Department of Management, University of Torino\\
Corso Unione Sovietica 218 bis, 10134 Torino, Italy\\
\texttt{luisa.tibiletti@unito.it}
\and
Simone Farinelli\\
Core Dynamics GmbH\\
Setzistrasse 7, 8272 Ermatingen, Switzerland\\
\texttt{simone@coredynamics.ch}
\and
Eric Dal Moro\\
Signal Iduna Reinsurance\\
Bundesplatz 1, 6300 Zug, Switzerland\\
\texttt{eric\_dal\_moro@yahoo.com}}

\date{}

\begin{document}
\maketitle

\begin{abstract}
Hierarchical copula models are widely used for aggregating market and credit risks across multi-level portfolio structures. Existing approaches often rely on Monte Carlo simulation, recursive sample reordering or high-dimensional numerical procedures. This paper develops an analytical framework for hierarchical copula aggregation based on explicit finite-sum representations of aggregate loss distributions and tail-risk measures for two-level market–credit aggregation with Gaussian and mirrored Clayton copulae under empirical leaf-level marginals.
We derive deterministic formulas for aggregate distribution functions, quantiles, expected tail losses and market–credit tail contributions, together with a compatibility condition linking local hierarchical copulae and global leaf-level dependence structures. Estimation-error expressions for quantiles and expected tail losses are also obtained in finite-sum form.
The paper further analyzes the qualitative asymptotic behaviour of hierarchical aggregation under increasing tree depth. Gaussian hierarchies are argued to remain diversification-dominated, whereas mirrored Clayton hierarchies propagate upper-tail dependence recursively and may amplify extreme losses across aggregation levels.
The resulting formulas provide a deterministic alternative to simulation-based aggregation methods and clarify how tail dependence propagates through hierarchical market--credit portfolios, with direct relevance for portfolio risk measurement and capital assessment.
\end{abstract}


\section{Introduction}

Aggregation of market and credit risk is a central problem in portfolio risk management, economic capital assessment and regulatory capital modelling. Institutions typically manage heterogeneous sources of risk separately, while capital and solvency assessment require the distribution of aggregate losses. Copula-based methods are attractive because they separate marginal modelling from dependence modelling, allowing different marginal risk models to be combined through an explicit dependence structure \citep{Sklar1959,Nelsen2006,McNeil2015}.
In practice, however, the calibration of copula parameters may itself
be challenging, particularly in insurance applications where tail
events are rare and historical observations are limited. Bayesian and
expert-based calibration approaches have therefore been proposed in the
actuarial literature \citep{ArbenzCanestraro2012}.
They have become a standard tool in financial
risk management \citep{Li2000,cherubini2004copula}.
For broad overviews of copula-based dependence modelling in risk
management and recent developments in copula theory, see
\citet{EmbrechtsMcNeilStraumann2002,Grober2022,Genest2024}.

Recent actuarial applications include reinsurance portfolio aggregation
and enterprise-wide solvency analysis, where dependence structures
critically influence diversification effects and aggregate capital
requirements \citep{Dias2025Extended}.

A major difficulty is that realistic portfolios are not flat. Market, credit and insurance risks are organized across books, portfolios, business lines and group-level entities. Hierarchical copula constructions address this issue by introducing dependence locally at different aggregation levels. Simulation-based approaches, including sample reordering methods, have been developed for hierarchical aggregation \citep{Arbenz2012}, while hierarchical Archimedean structures and related block constructions have been studied in actuarial and statistical settings \citep{Cossette2017,Chaoubi2021}. Vine and pair-copula constructions provide highly flexible high-dimensional alternatives \citep{Aas2009,Brechmann2013,czado2019vine,Nagler2021}, but their flexibility may come at the cost of economic interpretability and analytical tractability in capital aggregation problems. Recent discussions on simplified vine copula models and their practical limitations are given in \citet{Nagler2021}.

The present paper takes a complementary analytical perspective.
Rather than focusing on recursive simulation of hierarchical copula trees,
it derives explicit finite-sum representations for aggregate loss
distributions and tail-risk measures under hierarchical dependence.
The objective is therefore not the implementation of a simulation algorithm,
but the deterministic evaluation of quantiles, expected tail losses and
tail-risk contributions directly from empirical marginals and copula density
kernels.

The analytical framework also permits a structural comparison between
different dependence regimes. In particular, the paper highlights the
asymptotic distinction between Gaussian hierarchical aggregation, where
diversification effects remain asymptotically dominant, and mirrored
Clayton aggregation, where recursive upper-tail dependence may amplify
extreme losses across aggregation levels. The resulting formulas provide
an analytically tractable alternative to simulation-based hierarchical
aggregation procedures and clarify how tail dependence propagates through
multi-level portfolio structures.

The paper focuses on a two-level aggregation scheme with market and credit risk components. Market and credit losses are first aggregated within sub-portfolios, and sub-portfolio losses are then aggregated at the total portfolio level. For Gaussian and mirrored Clayton copulae, we derive finite-sum expressions for the distribution of the aggregate loss under empirical marginal distributions. We also derive formulas for quantiles, expected tail losses and market/credit tail contributions. These results are relevant for actuarial and financial applications where tail-sensitive capital measures are required and where repeated simulation of the copula may be computationally burdensome.
The asymptotic discussion further shows that hierarchical tail dependence
may fundamentally alter the scaling behaviour of high quantiles and
expected tail losses as the depth of the aggregation tree increases. The error analysis builds on standard quantile and expected tail loss estimation results \citep{Huber1981,OrdStuart1994,YamaiYoshiba2002,AcerbiTasche2002}. Moreover, expected tail loss contributions are closely related to coherent capital-allocation principles and Euler-type risk decompositions \citep{Tasche2008}.

The paper is organized as follows. Section~\ref{sec:framework}
introduces the hierarchical copula framework, the two-level
market--credit tree and a compatibility condition for hierarchical
copula specifications. Section~\ref{sec:copulas} recalls the Gaussian
and mirrored Clayton copulae. Section~\ref{sec:distribution} derives
explicit finite-sum aggregation formulas. Section~\ref{sec:tail} develops
deterministic formulas for quantiles, expected tail losses and tail
contributions. Section~\ref{sec:error} discusses estimation-error
expressions. Section~7 analyzes the asymptotic behaviour of hierarchical
aggregation under Gaussian and mirrored Clayton dependence. Section~\ref{sec:numerical}
provides a short numerical validation. Proofs are collected in
Appendix~\ref{app:proofs}.

\section{Hierarchical copula aggregation framework}\label{sec:framework}

Let $L_1$ and $L_2$ be loss random variables with distribution functions $F_1$ and $F_2$. By Sklar's theorem, a bivariate joint distribution can be represented as
\[
F_{12}(l_1,l_2)=C(F_1(l_1),F_2(l_2)),
\]
where $C$ is a copula. If $C$ is sufficiently smooth and $F_1,F_2$ admit densities $f_1,f_2$, the distribution function of the aggregate loss $L=L_1+L_2$ is
\begin{equation}\label{eq:general-aggregate}
F_L(l)=\int_{-\infty}^{+\infty}\int_{-\infty}^{l-l_2}
\partial_1\partial_2 C(F_1(l_1),F_2(l_2)) f_1(l_1)f_2(l_2)\dd l_1\dd l_2 .
\end{equation}
This formula is the basic analytical aggregation operator used throughout the paper.

Throughout the paper we adopt the convention
\[
\ThetaH(0)=1,
\]
so that finite weighted distributions are right-continuous cumulative
distribution functions.

\begin{figure}[ht]
\centering
\begin{forest}
for tree={
  draw,
  rounded corners,
  align=center,
  s sep=14mm,
  l sep=11mm,
  edge={-Latex},
  inner sep=2.5pt,
  font=\small
}
[$L$\\{\scriptsize total loss}
  [$L_1$\\{\scriptsize portfolio 1}
    [$L^{\MR}_1$\\{\scriptsize market risk}]
    [$L^{\CR}_1$\\{\scriptsize credit risk}]
  ]
  [$L_2$\\{\scriptsize portfolio 2}
    [$L^{\MR}_2$\\{\scriptsize market risk}]
    [$L^{\CR}_2$\\{\scriptsize credit risk}]
  ]
]
\end{forest}
\caption{Two-level hierarchical aggregation tree. In the Gaussian case, the node parameters are $(\gamma_1,\gamma_2,\rho)$, while in the mirrored Clayton case they are $(\zeta_1,\zeta_2,\theta)$. MR and CR denote market risk and credit risk, respectively.}
\label{fig:tree}
\end{figure}
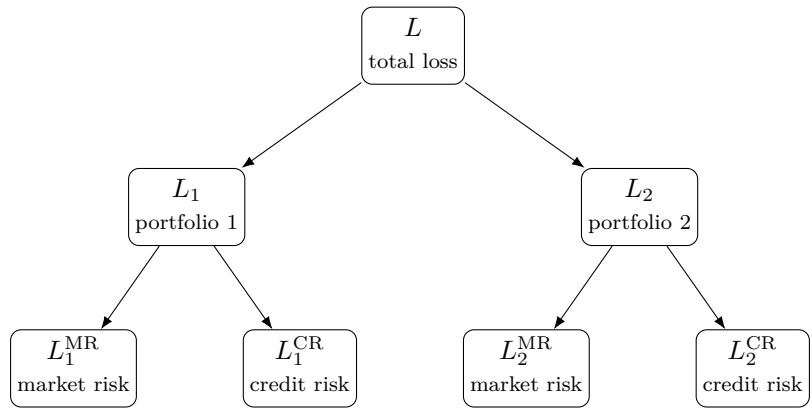

We consider a two-level portfolio tree. At the lowest level, two portfolios have market and credit risk components
\[
L_i=L_i^{\MR}+L_i^{\CR},\qquad i=1,2,
\]
and the total loss is
\[
L=L_1+L_2.
\]
The hierarchical structure is shown in Figure~\ref{fig:tree}. Copulae are specified at the market--credit aggregation level inside each portfolio and at the top aggregation level between the two portfolio losses.

The present approach is also related to the literature on analytical and
worst-case aggregation of dependent risks
\citep{puccetti2012bounds,EmbrechtsPuccetti2006}.

\subsection{Compatibility of local and global copulae}

The specification of copulae at several levels of a hierarchy is not purely formal. A global joint copula on the leaves must be compatible with the copulae specified at intermediate nodes and with the top-level copula used to aggregate the intermediate portfolio losses. The following theorem records the corresponding consistency condition in the two-level case of Figure~\ref{fig:tree}.

Let $C$ denote the four-dimensional copula of $(L_1^{\MR},L_1^{\CR},L_2^{\MR},L_2^{\CR})$, let $C^{1,\MR,\CR}$ and $C^{2,\MR,\CR}$ denote the within-portfolio copulae, and let $C^{1,2}$ denote the top-level copula of $(L_1,L_2)$. We write
\[
c=\partial_{(u_1^{\MR},u_1^{\CR},u_2^{\MR},u_2^{\CR})}C,
\qquad
c_i=\partial_{(u_i^{\MR},u_i^{\CR})}C^{i,\MR,\CR},
\qquad
c_{12}=\partial_{(u_1,u_2)}C^{1,2}.
\]
Let $F_i$ and $\rho_i$ denote the distribution function and density of the intermediate loss $L_i=L_i^{\MR}+L_i^{\CR}$ induced by the within-portfolio copula $C^{i,\MR,\CR}$.

\begin{theorem}[Hierarchical copula compatibility]\label{thm:compatibility}
Assume that all relevant distributions admit densities. A necessary condition for the existence of a global leaf-level copula $C$ compatible with the within-portfolio copulae and the top-level copula is that, for all $l_1,l_2\in\R$,
\begin{equation}\label{eq:compatibility}
\begin{split}
0=&\int_{\R^2}\rho_1^{\MR}(x_1)\rho_1^{\CR}(l_1-x_1)\rho_2^{\MR}(x_2)\rho_2^{\CR}(l_2-x_2)\\
&\quad\times\Big[
 c(u_1^{\MR},u_1^{\CR},u_2^{\MR},u_2^{\CR})
 -c_{12}(F_1(l_1),F_2(l_2))
  c_1(u_1^{\MR},u_1^{\CR})
  c_2(u_2^{\MR},u_2^{\CR})
\Big]\,\dd x_1\dd x_2,
\end{split}
\end{equation}
where
\[
(u_1^{\MR},u_1^{\CR},u_2^{\MR},u_2^{\CR})=
\big(F_1^{\MR}(x_1),F_1^{\CR}(l_1-x_1),F_2^{\MR}(x_2),F_2^{\CR}(l_2-x_2)\big).
\]
Moreover the global copula must satisfy the boundary conditions
\begin{equation}\label{eq:compatibility-boundary}
\begin{split}
\lim_{u_2^{\MR}\to1^-\!,\,u_2^{\CR}\to1^-}C(u_1^{\MR},u_1^{\CR},u_2^{\MR},u_2^{\CR})&=C^{1,\MR,\CR}(u_1^{\MR},u_1^{\CR}),\\
\lim_{u_1^{\MR}\to1^-\!,\,u_1^{\CR}\to1^-}C(u_1^{\MR},u_1^{\CR},u_2^{\MR},u_2^{\CR})&=C^{2,\MR,\CR}(u_2^{\MR},u_2^{\CR}).
\end{split}
\end{equation}
\end{theorem}

The theorem is not used as an algorithm. Its role is to clarify that a hierarchical specification imposes nontrivial consistency restrictions between local copulae, the top-level copula and the global leaf-level dependence structure. In particular, the top-level copula density $c_{12}$ enters explicitly in the consistency condition. The proof is given in Appendix~\ref{app:proofs}.

\section{Gaussian and mirrored Clayton copulae}\label{sec:copulas}

The Gaussian copula with correlation parameter $\rho\in(-1,1)$ is
\begin{equation}\label{eq:gaussian-copula}
C_\rho(u_1,u_2)=\Phi_\rho(\Phi^{-1}(u_1),\Phi^{-1}(u_2)),
\end{equation}
where $\Phi_\rho$ denotes the bivariate standard normal distribution function with correlation $\rho$. Its density is
\begin{equation}\label{eq:gaussian-density-kernel}
c_\rho(u_1,u_2)=\frac{1}{\sqrt{1-\rho^2}}
\exp\left\{-\frac{\rho^2}{2(1-\rho^2)}\left(z_1^2-\frac{2}{\rho}z_1z_2+z_2^2\right)\right\},
\end{equation}
with $z_i=\Phi^{-1}(u_i)$.

The mirrored, or survival, Clayton copula with parameter $\theta>0$ is
\begin{equation}\label{eq:survival-clayton}
C_\theta(u_1,u_2)=u_1+u_2-1+\big((1-u_1)^{-\theta}+(1-u_2)^{-\theta}-1\big)^{-1/\theta}.
\end{equation}
Its density is
\begin{equation}\label{eq:survival-clayton-density}
c_\theta(u_1,u_2)=(1+\theta)\big((1-u_1)^{-\theta}+(1-u_2)^{-\theta}-1\big)^{-1/\theta-2}
\big((1-u_1)(1-u_2)\big)^{-\theta-1}.
\end{equation}
The mirrored Clayton family is useful for modeling upper-tail dependence in loss variables.

For these copulae, Kendall's tau is related to the copula parameters by
\begin{equation}\label{eq:kendall-maps}
\tau_{\Gauss}=\frac{2}{\pi}\arcsin(\rho),\qquad
\tau_{\Clay}=\frac{\theta}{\theta+2}.
\end{equation}
The first identity follows from the elliptical structure of the Gaussian copula \citep{Fang2002}; the second follows by direct integration of the copula density.

\section{Explicit finite-sum aggregate loss distributions}\label{sec:distribution}

At the lowest level, the marginal distributions are represented by empirical distributions. Let
\begin{equation}\label{eq:empirical-marginals}
F_r(l)=\frac{1}{n}\sum_{m=1}^n \ThetaH(l-x_r^m),
\qquad
F'_r(l)=\frac{1}{n}\sum_{m=1}^n \delta(l-x_r^m),
\end{equation}
where $r$ indexes the relevant market or credit leaf, $\ThetaH$ denotes the Heaviside function and $\delta$ denotes the Dirac delta.

\begin{remark}[Quadrature consistency]
The finite-sum constructions may be interpreted either as deterministic quadrature rules on empirical grids or as weighted empirical approximations generated from sampled leaf marginals. Under standard convergence of the empirical marginals and continuity of the copula kernels away from endpoint singularities, the normalized finite sums converge formally to the corresponding continuous aggregation integrals as $n\to\infty$.
\end{remark}

The use of empirical marginals should be understood as a deterministic finite-sample discretization of the copula-integral representation, rather than as the assertion that the empirical distribution determines a unique discrete copula in the strict Sklar-theoretic sense. The copula density kernels provide the quadrature weights used to evaluate the continuous aggregation operator on the empirical grid. Substituting empirical marginals into the aggregation operator \eqref{eq:general-aggregate} transforms the integral representation into finite weighted sums. This is the basic mechanism behind the following propositions.

\subsection{Two levels of Gaussian copulae}

\begin{figure}[ht]
\centering
\begin{forest}
for tree={
  draw,
  rounded corners,
  align=center,
  s sep=14mm,
  l sep=11mm,
  edge={-Latex},
  inner sep=2.5pt,
  font=\small
}
[$L$\\{\scriptsize top node}\\{\scriptsize Gaussian copula $\rho$}
  [$L_1$\\{\scriptsize portfolio 1}\\{\scriptsize Gaussian copula $\gamma_1$}
    [$L^{\MR}_1$\\{\scriptsize market risk}]
    [$L^{\CR}_1$\\{\scriptsize credit risk}]
  ]
  [$L_2$\\{\scriptsize portfolio 2}\\{\scriptsize Gaussian copula $\gamma_2$}
    [$L^{\MR}_2$\\{\scriptsize market risk}]
    [$L^{\CR}_2$\\{\scriptsize credit risk}]
  ]
]
\end{forest}
\caption{Gaussian hierarchical aggregation structure. The parameters
$\gamma_1$ and $\gamma_2$ govern the within-portfolio market--credit
dependence, while $\rho$ governs the top-level dependence between
portfolio losses $L_1$ and $L_2$.}
\label{fig:gaussian-tree}
\end{figure}
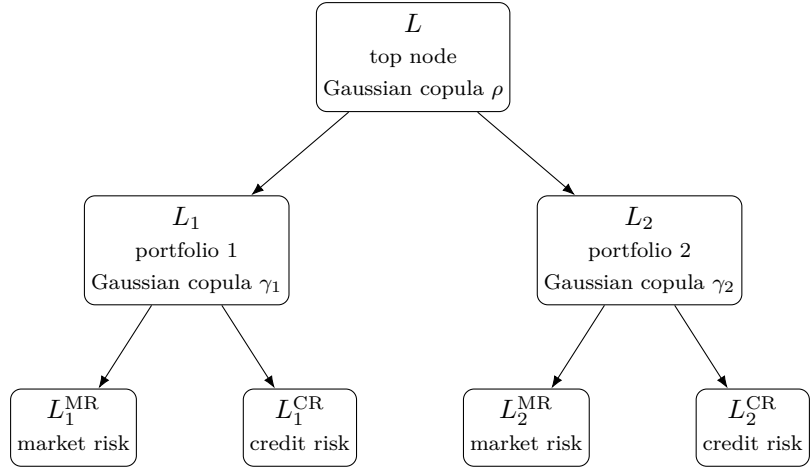

The parameterized Gaussian hierarchy is displayed in Figure~\ref{fig:gaussian-tree}. Let $\gamma_1$ and $\gamma_2$ denote the Gaussian copula correlations between market and credit losses inside portfolios $1$ and $2$, respectively. Let $\rho$ denote the Gaussian copula correlation between $L_1$ and $L_2$ at the top level. We write $u_m=(m-0.5)/n,\qquad z_m=\Phi^{-1}(u_m)$,
\[
Z_1(j,i)=\Phi^{-1}\big(F_1(x_{1j}^{\MR}+x_{1i}^{\CR})\big),\qquad
Z_2(h,k)=\Phi^{-1}\big(F_2(x_{2h}^{\MR}+x_{2k}^{\CR})\big).
\]

\begin{proposition}[Gaussian hierarchical aggregate distribution]\label{prop:gaussian-total}
Under empirical leaf marginals and Gaussian copulae at both aggregation levels, the total loss distribution admits the finite-sum representation
\begin{equation}\label{eq:gaussian-total}
F_L(l)=\frac{1}{n^4\sqrt{(1-\rho^2)(1-\gamma_1^2)(1-\gamma_2^2)}}
\sum_{j,i,h,k=1}^n \ThetaH\big(l-x_{1j}^{\MR}-x_{1i}^{\CR}-x_{2h}^{\MR}-x_{2k}^{\CR}\big) W^{G}_{j i h k},
\end{equation}
where
\begin{align}\label{eq:gaussian-weight}
W^{G}_{j i h k}
&=\exp\Bigg[-\frac{\gamma_1^2}{2(1-\gamma_1^2)}
\left(z_j^2-\frac{2}{\gamma_1}z_jz_i+z_i^2\right)
-\frac{\gamma_2^2}{2(1-\gamma_2^2)}
\left(z_h^2-\frac{2}{\gamma_2}z_hz_k+z_k^2\right)\notag\\
&\qquad
-\frac{\rho^2}{2(1-\rho^2)}
\left( Z_1(j,i)^2-\frac{2}{\rho}Z_1(j,i)Z_2(h,k)+Z_2(h,k)^2\right)
\Bigg].
\end{align}
\end{proposition}

\subsection{Two levels of mirrored Clayton copulae}

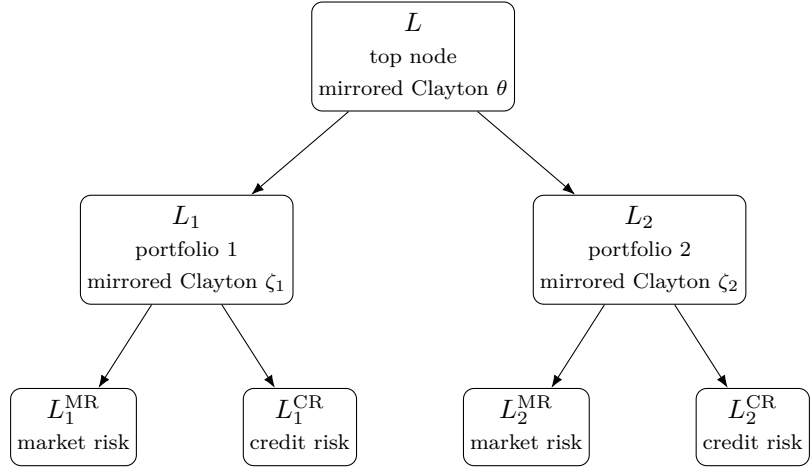
\begin{figure}[ht]
\centering
\begin{forest}
for tree={
  draw,
  rounded corners,
  align=center,
  s sep=14mm,
  l sep=11mm,
  edge={-Latex},
  inner sep=2.5pt,
  font=\small
}
[$L$\\{\scriptsize top node}\\{\scriptsize mirrored Clayton $\theta$}
  [$L_1$\\{\scriptsize portfolio 1}\\{\scriptsize mirrored Clayton $\zeta_1$}
    [$L^{\MR}_1$\\{\scriptsize market risk}]
    [$L^{\CR}_1$\\{\scriptsize credit risk}]
  ]
  [$L_2$\\{\scriptsize portfolio 2}\\{\scriptsize mirrored Clayton $\zeta_2$}
    [$L^{\MR}_2$\\{\scriptsize market risk}]
    [$L^{\CR}_2$\\{\scriptsize credit risk}]
  ]
]
\end{forest}
\caption{Mirrored Clayton hierarchical aggregation structure. The
parameters $\zeta_1$ and $\zeta_2$ govern the within-portfolio
upper-tail dependence between market and credit losses, while $\theta$
governs the top-level upper-tail dependence between portfolio losses
$L_1$ and $L_2$.}
\label{fig:clayton-tree}
\end{figure}

The parameterized mirrored Clayton hierarchy is displayed in Figure~\ref{fig:clayton-tree}. Let $\zeta_1$ and $\zeta_2$ denote the mirrored Clayton parameters inside portfolios $1$ and $2$, and let $\theta$ denote the top-level mirrored Clayton parameter. Set $u_m=(m-0.5)/n$.

\begin{proposition}[Mirrored Clayton hierarchical aggregate distribution]\label{prop:clayton-total}
Under empirical leaf marginals and mirrored Clayton copulae at both aggregation levels, the total loss distribution admits the representation
\begin{equation}\label{eq:clayton-total}
F_L(l)=\frac{(1+\theta)(1+\zeta_1)(1+\zeta_2)}{n^4}
\sum_{j,i,h,k=1}^n \ThetaH\big(l-x_{1j}^{\MR}-x_{1i}^{\CR}-x_{2h}^{\MR}-x_{2k}^{\CR}\big) W^{C}_{j i h k},
\end{equation}
where
\begin{align}\label{eq:clayton-weight}
W^{C}_{j i h k}
&=\Big[ (1-u_i)^{-\zeta_1}+(1-u_j)^{-\zeta_1}-1\Big]^{-1/\zeta_1-2}
\big[(1-u_i)(1-u_j)\big]^{-\zeta_1-1}\notag\\
&\quad\times
\Big[ (1-u_k)^{-\zeta_2}+(1-u_h)^{-\zeta_2}-1\Big]^{-1/\zeta_2-2}
\big[(1-u_k)(1-u_h)\big]^{-\zeta_2-1}\notag\\
&\quad\times
\Big[ (1-F_1(x_{1j}^{\MR}+x_{1i}^{\CR}))^{-\theta}+(1-F_2(x_{2h}^{\MR}+x_{2k}^{\CR}))^{-\theta}-1\Big]^{-1/\theta-2}\notag\\
&\quad\times
\Big[(1-F_1(x_{1j}^{\MR}+x_{1i}^{\CR}))(1-F_2(x_{2h}^{\MR}+x_{2k}^{\CR}))\Big]^{-\theta-1}.
\end{align}
\end{proposition}

\begin{remark}[Normalization]
For each fixed hierarchy and copula-parameter specification, the coefficients in the finite-sum distributions are interpreted as quadrature weights induced by copula density kernels on the empirical grid. They are nonnegative by construction. In implementation they are normalized so that
\[
\sum_{j,i,h,k} p_{j i h k}=1,
\]
and hence the resulting finite-sum distribution is a probability distribution. In exact arithmetic this normalization is the discrete counterpart of integrating the copula density over the unit square at each node.
\end{remark}

\begin{remark}
Equations \eqref{eq:gaussian-total} and \eqref{eq:clayton-total} are not simulation algorithms. They define weighted empirical distributions of the total loss induced by the hierarchical copula specification. The resulting distribution can be evaluated deterministically at any loss level $l$.
\end{remark}

\section{Explicit finite-sum quantiles, expected tail losses and tail contributions}\label{sec:tail}

Let $F_L$ be a finite weighted distribution of the form
\begin{equation}\label{eq:generic-weighted-distribution}
F_L(l)=\sum_{r\in\mathcal R} p_r\,\ThetaH(l-\ell_r),
\qquad p_r\ge0,
\qquad \sum_{r\in\mathcal R}p_r=1,
\end{equation}
where $\ell_r$ denotes an aggregate loss realization and $p_r$ is its copula-induced weight.

\begin{definition}
For $\alpha\in(0,1)$, the lower $\alpha$-quantile of $L$ is
\begin{equation}
q_\alpha(L)=\inf\{l:F_L(l)\ge\alpha\}.
\end{equation}
For a finite weighted distribution, let
\[
q=q_\alpha(L),\qquad
M_>(q)=\sum_{r\in\mathcal R}p_r\,\1_{\{\ell_r>q\}},\qquad
M_=(q)=\sum_{r\in\mathcal R}p_r\,\1_{\{\ell_r=q\}}.
\]
The amount of mass at the quantile needed to complete the upper tail of probability $1-\alpha$ is
\[
\eta_\alpha(q)=(1-\alpha)-M_>(q),\qquad 0\le \eta_\alpha(q)\le M_=(q).
\]
The expected tail loss at confidence level $\alpha$ is then defined by
\begin{equation}\label{eq:etl-discrete-atom}
\ETL_\alpha(L)=\frac{1}{1-\alpha}
\left(
\sum_{r\in\mathcal R}p_r\ell_r\,\1_{\{\ell_r>q\}}
+\eta_\alpha(q)q
\right).
\end{equation}
This is the standard finite-distribution adjustment at an atom of the quantile. When no atom correction is needed, \eqref{eq:etl-discrete-atom} reduces to the strict-tail expression.
\end{definition}

\begin{proposition}[Deterministic quantile extraction]\label{prop:var-finite}
Let the losses $\ell_r$ in \eqref{eq:generic-weighted-distribution} be ordered increasingly as $\ell_{(1)}\le\cdots\le\ell_{(N)}$, with corresponding weights $p_{(m)}$. Then
\begin{equation}\label{eq:var-finite}
q_\alpha(L)=\ell_{(m_\alpha)},\qquad
m_\alpha=\min\left\{m:\sum_{q=1}^m p_{(q)}\ge\alpha\right\}.
\end{equation}
\end{proposition}

\begin{proposition}[Gaussian market--credit tail contributions]\label{prop:gaussian-tail}
Assume that total market and credit losses $L^{\MR}$ and $L^{\CR}$ have empirical marginals and are joined by a Gaussian copula with parameter $\gamma$. Set $z_i=\Phi^{-1}(i/n)$ and define the Gaussian rank weight
\begin{equation}\label{eq:gaussian-tail-weight}
A_{ij}^{\gamma}:=\exp\left[-\frac{\gamma^2}{2(1-\gamma^2)}(z_i-z_j)^2\right].
\end{equation}
Then
\begin{equation}\label{eq:gaussian-tail-contrib}
\begin{split}
\ETLCont_\alpha(L^{\MR},L)&=\frac{1}{(1-\alpha)n^2\sqrt{1-\gamma^2}}
\sum_{i,j=1}^n\ThetaH(l^{\MR,i}+l^{\CR,j}-q_\alpha(L))\,l^{\MR,i}\,A_{ij}^{\gamma},\\
\ETLCont_\alpha(L^{\CR},L)&=\frac{1}{(1-\alpha)n^2\sqrt{1-\gamma^2}}
\sum_{i,j=1}^n\ThetaH(l^{\MR,i}+l^{\CR,j}-q_\alpha(L))\,l^{\CR,j}\,A_{ij}^{\gamma}.
\end{split}
\end{equation}
The expected tail loss is the sum of the two tail contributions.

If the aggregate distribution possesses an atom at $q_\alpha(L)$,
the strict-tail indicator in \eqref{eq:gaussian-tail-contrib}
is replaced by the atom-corrected finite-distribution convention
\eqref{eq:general-etl-atom-market}--\eqref{eq:general-etl-atom-credit},
with weights induced by the Gaussian copula kernel.
\end{proposition}

\begin{proposition}[Mirrored Clayton market--credit tail contributions]\label{prop:clayton-tail}
Assume that total market and credit losses $L^{\MR}$ and $L^{\CR}$ have empirical marginals and are joined by a mirrored Clayton copula with parameter $\zeta$. Then
\begin{equation}\label{eq:clayton-tail-contrib}
\begin{split}
\ETLCont_\alpha(L^{\MR},L)&=\frac{1+\zeta}{(1-\alpha)n^2}
\sum_{i,j=1}^n\ThetaH(l^{\MR,i}+l^{\CR,j}-q_\alpha(L))\,l^{\MR,i}\,A_{ij}^{\zeta},\\
\ETLCont_\alpha(L^{\CR},L)&=\frac{1+\zeta}{(1-\alpha)n^2}
\sum_{i,j=1}^n\ThetaH(l^{\MR,i}+l^{\CR,j}-q_\alpha(L))\,l^{\CR,j}\,A_{ij}^{\zeta},
\end{split}
\end{equation}
where
\begin{equation}
A_{ij}^{\zeta}=\left[\left(1-\frac{i}{n}\right)^{-\zeta}+\left(1-\frac{j}{n}\right)^{-\zeta}-1\right]^{-1/\zeta-2}
\left[\left(1-\frac{i}{n}\right)\left(1-\frac{j}{n}\right)\right]^{-\zeta-1}.
\end{equation}
The expected tail loss is the sum of the two tail contributions.

If the aggregate distribution possesses an atom at $q_\alpha(L)$,
the strict-tail indicator in \eqref{eq:clayton-tail-contrib}
is replaced by the atom-corrected finite-distribution convention
\eqref{eq:general-etl-atom-market}--\eqref{eq:general-etl-atom-credit},
with weights induced by the mirrored Clayton copula kernel.
\end{proposition}

\begin{remark}[Atom correction for tail contributions]
Suppose that the aggregate loss distribution has an atom at
$q_\alpha(L)$. Let
\[
A_q:=\{r\in\mathcal R:\ell_r=q_\alpha(L)\},
\qquad
M_{=}(q):=\sum_{r\in A_q}p_r.
\]
Then the expected tail loss contributions are defined by
\begin{equation}\label{eq:general-etl-atom-market}
\ETLCont_\alpha(L^{\MR},L)
=
\frac{1}{1-\alpha}
\left(
\sum_{\ell_r>q}p_r\,\ell_r^{\MR}
+
\eta_\alpha(q)
\sum_{r\in A_q}
\frac{p_r}{M_{=}(q)}
\,\ell_r^{\MR}
\right),
\end{equation}
and analogously
\begin{equation}\label{eq:general-etl-atom-credit}
\ETLCont_\alpha(L^{\CR},L)
=
\frac{1}{1-\alpha}
\left(
\sum_{\ell_r>q}p_r\,\ell_r^{\CR}
+
\eta_\alpha(q)
\sum_{r\in A_q}
\frac{p_r}{M_{=}(q)}
\,\ell_r^{\CR}
\right),
\end{equation}
With this convention,
\[
\ETL_\alpha(L)
=
\ETLCont_\alpha(L^{\MR},L)
+
\ETLCont_\alpha(L^{\CR},L).
\]
The quantile atom is therefore shared proportionally according to the
copula-weighted contributions of the scenarios belonging to the atom.
\end{remark}

\begin{remark}
The normalizing factor is written in the standard expected tail loss convention for the upper tail, namely $1-\alpha$. If one denotes the tail probability itself by $\alpha$, the same expressions are obtained by replacing $1-\alpha$ with $\alpha$.
\end{remark}

\section{Estimation error for quantiles and expected tail losses}\label{sec:error}

The finite-sum representations above eliminate copula simulation error once the marginal samples and copula parameters are fixed. In practice, the empirical inputs are themselves estimated. We recall a univariate benchmark for quantile and expected tail loss estimation error, following \citet{YamaiYoshiba2002}.

\begin{proposition}[Quantile and ETL estimation error]\label{prop:error}
Let $X$ be a univariate loss with distribution function $F$ and density $F'$. Let $X_1,\dots,X_{n-1}$ be ordered in decreasing order. The estimators
\[
\widehat q_\alpha(X)=X_{[n(1-\alpha)+1]},\qquad
\widehat{\ETL}_\alpha(X)=\frac{X_1+\cdots+X_{[n(1-\alpha)+1]}}{n(1-\alpha)+1}
\]
have standard deviations of the form
\begin{equation}\label{eq:q-error}
\sigma(\widehat q_\alpha(X))=\sqrt{\frac{\alpha(1-\alpha)}{n}}\frac{1}{F'(\widehat q_\alpha(X))},
\end{equation}
and
\begin{equation}\label{eq:etl-error}
\begin{split}
\sigma(\widehat{\ETL}_\alpha(X))=\frac{1}{\sqrt n}\Bigg\{&
\frac{\alpha \widehat q_\alpha(X)^2+(1-\beta)\widehat q_\beta(X)^2}{(\alpha-\beta)^2}
+\frac{1}{(\alpha-\beta)^2}\int_{\widehat q_\alpha(X)}^{\widehat q_\beta(X)}x^2F'(x)\dd x\\
&-\frac{1}{(\alpha-\beta)^2}\left[(1-\beta)\widehat q_\beta(X)+\alpha\widehat q_\alpha(X)
+\int_{\widehat q_\alpha(X)}^{\widehat q_\beta(X)}xF'(x)\dd x\right]^2
\Bigg\}^{1/2},
\end{split}
\end{equation}
for an auxiliary confidence level $\beta$.
\end{proposition}

In the present setting, $F'$ is a weighted sum of Dirac measures obtained from the corresponding explicit finite-sum distribution in Section~\ref{sec:distribution}. The following corollary records the resulting finite-sum form.

\begin{corollary}[Finite-sum estimation-error formulas]\label{cor:finite-error}
Let
\begin{equation}\label{eq:discrete-density}
F_L(l)=\sum_{r\in\mathcal R}p_r\ThetaH(l-\ell_r),\qquad
F'_L(l)=\sum_{r\in\mathcal R}p_r\delta(l-\ell_r),
\end{equation}
where the weights $p_r$ are obtained from the Gaussian or mirrored Clayton finite-sum representations in Section~\ref{sec:distribution}. In the absence of atoms at the relevant quantile endpoints, the integral terms in \eqref{eq:etl-error} reduce to the Dirac sums
\begin{equation}\label{eq:dirac-moment-sums-simple}
\int_{\widehat q_\alpha}^{\widehat q_\beta} x^mF'_L(x)\dd x
=\sum_{r\in\mathcal R}p_r\ell_r^m\,
\1_{\{\widehat q_\alpha\le \ell_r\le \widehat q_\beta\}},
\qquad m=1,2.
\end{equation}
For a genuinely finite distribution, atoms at the quantile levels should be treated with the same convention as in Definition~1. For $a\in(0,1)$, set $q_a=q_a(L)$ and define the atom-corrected upper-tail moments
\begin{equation}\label{eq:atom-corrected-upper-moments}
\mathcal U_m(a)
:=
\sum_{r\in\mathcal R}p_r\ell_r^m\,\1_{\{\ell_r>q_a\}}
+
\eta_a(q_a)q_a^m,
\qquad m=1,2.
\end{equation}
For $\beta>\alpha$, the atom-corrected finite-distribution analogue of the moment over the interval between the two quantiles is
\begin{equation}\label{eq:atom-corrected-interval-moments}
\mathcal I_m(\alpha,\beta)
:=
\mathcal U_m(\alpha)-\mathcal U_m(\beta),
\qquad m=1,2.
\end{equation}
Equivalently, \eqref{eq:atom-corrected-interval-moments} includes the full mass of the scenarios strictly between the two quantiles and only the fractions of the endpoint atoms required by the finite-distribution ETL convention.
Consequently, the atom-corrected finite-sum version of \eqref{eq:etl-error} is
\begin{equation}\label{eq:finite-etl-error}
\begin{split}
\sigma(\widehat{\ETL}_\alpha(L))=\frac{1}{\sqrt n}\Bigg\{&
\frac{\alpha\widehat q_\alpha^2+(1-\beta)\widehat q_\beta^2}{(\alpha-\beta)^2}
+\frac{\mathcal I_2(\alpha,\beta)}{(\alpha-\beta)^2}\\
&-\frac{1}{(\alpha-\beta)^2}
\left[(1-\beta)\widehat q_\beta+\alpha\widehat q_\alpha
+\mathcal I_1(\alpha,\beta)\right]^2
\Bigg\}^{1/2}.
\end{split}
\end{equation}
If there are no endpoint atoms, \eqref{eq:finite-etl-error} reduces to the formula obtained by substituting the ordinary Dirac sums \eqref{eq:dirac-moment-sums-simple} into \eqref{eq:etl-error}.
For the quantile standard deviation, the point density $F'_L(\widehat q_\alpha)$ is a Dirac mass. In numerical use one replaces it by a local effective density, for example
\begin{equation}\label{eq:local-density}
f_\Delta(\widehat q_\alpha):=\frac{F_L(\widehat q_\alpha+\Delta)-F_L(\widehat q_\alpha-\Delta)}{2\Delta}
=\frac{1}{2\Delta}\sum_{r\in\mathcal R}p_r\,
\1_{\{|\ell_r-\widehat q_\alpha|\le\Delta\}},
\end{equation}
which gives the finite-sum approximation
\begin{equation}\label{eq:finite-q-error}
\sigma(\widehat q_\alpha(L))\approx
\sqrt{\frac{\alpha(1-\alpha)}{n}}\,
\frac{1}{f_\Delta(\widehat q_\alpha)}.
\end{equation}
\end{corollary}

For the Gaussian hierarchy, one takes $r=(j,i,h,k)$,
\begin{equation}
\ell_{j i h k}=x_{1j}^{\MR}+x_{1i}^{\CR}+x_{2h}^{\MR}+x_{2k}^{\CR},\qquad
p_{j i h k}^{G}=\frac{W^G_{j i h k}}{n^4\sqrt{(1-\rho^2)(1-\gamma_1^2)(1-\gamma_2^2)}}.
\end{equation}
For the mirrored Clayton hierarchy, one takes the same loss grid and
\begin{equation}
p_{j i h k}^{C}=\frac{(1+\theta)(1+\zeta_1)(1+\zeta_2)}{n^4}W^C_{j i h k}.
\end{equation}
Substituting $p^G$ or $p^C$ into \eqref{eq:finite-etl-error}--\eqref{eq:finite-q-error} gives the desired closed finite sums for the standard deviation of the quantile and of the expected tail loss.

\section{Qualitative asymptotic behaviour under increasing hierarchical depth}
\label{sec:asymptotic}

We now discuss the qualitative asymptotic behaviour of the top-level quantile and expected tail loss when the number of aggregation levels increases. The purpose of this section is not to prove a universal limit theorem for arbitrary hierarchical copula trees. Rather, it identifies the different mechanisms by which Gaussian and upper-tail-dependent copulae affect tail aggregation in deep hierarchical structures.

The following discussion combines classical results on tail dependence and heavy-tail asymptotics for copulas with the recursive structure of hierarchical aggregation trees. While asymptotic tail dependence properties of Gaussian and Archimedean copulas are well known \citep{embrechts2001modelling,McNeil2015,Joe2014}, their propagation through recursive hierarchical aggregation structures appears to be less explored.

Analytical approximations for aggregate portfolio losses also play a central role in asymptotic risk-factor models used in banking regulation \citep{gordy2003risk}. Related asymptotic aggregation questions arise in insurance solvency frameworks and enterprise capital models under Solvency II \citep{SolvencyII2009}.

Consider a balanced aggregation tree with branching number $b$ and depth $m$. The number of leaves is
\[
N=b^m.
\]
Assume for simplicity that all leaf losses are identically distributed, all internal nodes use the same copula family, and copula parameters are homogeneous across the tree. The aggregate portfolio loss at the top level is denoted by
\[
L_N=\sum_{i=1}^N X_i.
\]

\subsection{Gaussian hierarchical aggregation}

For Gaussian copulas with correlation parameter $\rho<1$, the upper-tail dependence coefficient vanishes \citep{McNeil2015}:
\[
\lambda_U^{\mathrm{Gauss}}=0.
\]
Thus, although dependence affects moderate losses and the variance of the aggregate portfolio, extreme losses do not remain asymptotically clustered in the far tail.

Assume that the leaf marginals possess finite variance and denote
\[
\mu=\mathbb{E}[X_i],
\qquad
\sigma^2=\mathrm{Var}(X_i).
\]
For any covariance structure generated by the hierarchy,
\[
\mathrm{Var}(L_N)=N\sigma^2+2\sum_{i<j}\mathrm{Cov}(X_i,X_j).
\]
Consequently a Gaussian-type approximation, when justified by the effective dependence structure, gives
\[
q_\alpha^{\mathrm{Gauss}}(L_N)
\approx
N\mu+z_\alpha\sqrt{\mathrm{Var}(L_N)},
\qquad
z_\alpha=\Phi^{-1}(\alpha),
\]
and
\[
\mathrm{ETL}_\alpha^{\mathrm{Gauss}}(L_N)
\approx
N\mu+
\sqrt{\mathrm{Var}(L_N)}\frac{\varphi(z_\alpha)}{1-\alpha}.
\]
If average correlations remain bounded away from zero, $\mathrm{Var}(L_N)$ may grow of order $N^2$ rather than $N$. Thus the term ``diversification'' should not be read as a universal $\sqrt N$ scaling statement. The relevant point for the present comparison is instead that Gaussian copulae do not generate recursive upper-tail clustering: their effect is transmitted through covariance rather than through a positive upper-tail dependence coefficient.

\subsection{Mirrored Clayton hierarchical aggregation}

The qualitative regime is different for mirrored Clayton copulae. In this case the upper-tail dependence coefficient is strictly positive \citep{embrechts2001modelling,Joe2014}:
\[
\lambda_U^{\mathrm{Clayton}}=2^{-1/\theta}.
\]
Consequently, large losses may remain clustered throughout the hierarchy. This effect becomes particularly pronounced when the leaf marginals are heavy-tailed.

To avoid confusion with the extreme-value shape parameter, we denote the Pareto tail exponent by $\beta>0$ and assume that the leaf losses possess regularly varying tails
\[
\mathbb{P}(X_i>x)\sim Cx^{-\beta},
\qquad x\to\infty.
\]
For a fixed depth $m$, a useful heuristic representation of the aggregate upper tail is
\[
\mathbb{P}(L_N>x)
\sim
\kappa_m\,\mathbb{P}(X>x),
\]
where $\kappa_m$ is an effective tail-clustering coefficient induced by the hierarchical dependence structure. In a homogeneous tree, $\kappa_m$ summarizes how upper-tail dependence is transmitted through the successive aggregation nodes. It depends on the branching number, the local copula parameters and the local coefficient $\lambda_U$.

One may interpret $\kappa_m$ recursively as the multiplicative tail-equivalent coefficient generated by replacing each internal aggregation node by its tail-equivalent contribution. The precise recursion is model-specific; for the purposes of this paper the essential point is that $\kappa_m$ can increase with the depth of a homogeneous upper-tail-dependent hierarchy, whereas no analogous upper-tail clustering coefficient is present in the Gaussian case.

The corresponding high-quantile approximation is
\[
q_\alpha^{\mathrm{Clayton}}(L_N)
\sim
F_X^{-1}\!\left(
1-\frac{1-\alpha}{\kappa_m}
\right).
\]
If the tail exponent satisfies $\beta>1$, so that the mean excess is finite, the expected tail loss satisfies the usual regularly varying approximation
\[
\mathrm{ETL}_\alpha^{\mathrm{Clayton}}(L_N)
\sim
\frac{\beta}{\beta-1}\,
q_\alpha^{\mathrm{Clayton}}(L_N).
\]
Equivalently, in extreme-value notation with shape parameter $\xi=1/\beta<1$, this factor is $1/(1-\xi)$.

Thus, unlike the Gaussian case, the mirrored Clayton hierarchy may amplify extreme losses through recursive tail clustering. As the number of aggregation levels increases, the top-level quantiles and expected tail losses may grow substantially faster than under a Gaussian dependence structure with comparable rank correlations.

This recursive amplification mechanism may become particularly relevant for hierarchical internal models in banking and (re)insurance, where deep aggregation trees combined with tail-dependent copulae may substantially increase regulatory capital requirements.

\subsection{Interpretation}

The previous discussion highlights a fundamental distinction between Gaussian and tail-dependent hierarchical copula structures. In the Gaussian case, increasing the number of aggregation levels mainly changes the covariance structure of the aggregate portfolio. Depending on the average correlation, the variance may scale between order $N$ and order $N^2$, but there is no recursive upper-tail dependence mechanism.

By contrast, mirrored Clayton copulae propagate tail dependence through the hierarchy. Extreme losses generated at lower aggregation levels may therefore remain synchronized at higher levels, producing stronger tail amplification and significantly larger expected tail losses.

For illustration, consider a homogeneous Gaussian hierarchy with pairwise effective correlation $\bar\rho_N=0.5$. In this case,
\[
\mathrm{Var}(L_N)=N\sigma^2\bigl(1+0.5(N-1)\bigr),
\]
so that the portfolio standard deviation grows proportionally to $N$. This example shows that Gaussian dependence does not always imply classical $\sqrt N$ diversification. Nevertheless, the corresponding quantile and expected tail loss are still governed by Gaussian covariance scaling rather than by recursive upper-tail clustering. This behaviour contrasts with the amplification mechanism observed under mirrored Clayton dependence.

This effect is generally more regular for expected tail loss than for quantiles, since expected tail loss is subadditive whereas quantiles need not be. Consequently, modifications of the hierarchical structure may affect high quantiles in a non-monotone manner.

Overall, the asymptotic distinction becomes increasingly important for deep hierarchical structures and high confidence levels, where tail dependence dominates the behaviour of coherent risk measures.

\section{Numerical validation}\label{sec:numerical}

We retain the illustrative two-level structure of Figure~\ref{fig:tree}. Market-risk marginals are represented by lognormal distributions with parameters $\mu_M=0$ and $\sigma_M=0.25$, while credit-risk marginals are modeled by shifted Pareto distributions with shift parameter $x_0=1$, scale parameter $c=1$ and tail exponent $\beta=2.5$. The Gaussian node parameters are chosen as $\gamma_1=0.75$, $\gamma_2=0.5$ and $\rho=0.5$. The mirrored Clayton parameters are chosen as $\zeta_1=2.3468$, $\zeta_2=1.0$ and $\theta=1.0$.

The empirical marginals are generated from samples of size $n=200$ at each leaf.
Monte Carlo benchmark distributions are obtained from $10^6$ simulated hierarchical copula draws using the same plotting-position convention $u_m=(m-0.5)/n$. Before normalization, the Gaussian and mirrored Clayton weight sums differed from one by less than $10^{-4}$ in all reported calculations.

In addition to the pointwise CDF comparisons reported below, the corresponding $99\%$ quantile and expected tail loss discrepancies remained below $1.5\%$ relative error across the validation grid. The maximum absolute CDF discrepancy observed on the evaluation grid was below $0.02$ for both copula families.

The purpose of the numerical section is not to introduce a new simulation method, but to validate that the analytical finite-sum formulas reproduce benchmark distributions obtained from direct copula simulation. The formula values are obtained by deterministic summation over the empirical grid; the simulation values are Monte Carlo benchmarks generated from the corresponding hierarchical copula specification. Table~\ref{tab:validation} reports representative distribution values at the second aggregation level.

\begin{table}[ht]
\centering
\caption{Representative validation of finite-sum formulas against simulation benchmarks.}
\label{tab:validation}
\begin{tabular}{lccc}
\toprule
Copula family & Loss level & Simulation CDF & Formula CDF \\
\midrule
Gaussian & 25 & 43.1\% & 42.4\% \\
Gaussian & 28 & 60.1\% & 59.5\% \\
Gaussian & 30 & 67.8\% & 67.7\% \\
Mirrored Clayton & 25 & 47.3\% & 45.8\% \\
Mirrored Clayton & 28 & 63.9\% & 62.6\% \\
Mirrored Clayton & 30 & 71.5\% & 70.1\% \\
\bottomrule
\end{tabular}
\end{table}

The agreement between the simulation benchmark and the finite-sum formulas supports the analytical derivation. The residual differences are attributable to finite discretization and simulation variability. The finite-sum probabilities are normalized before computing distribution values, which also provides a numerical check that the copula-induced weights define a probability distribution.

The formulas make clear why computational cost grows rapidly with the number of leaf marginals: a two-level four-leaf construction requires summation over $n^4$ combinations. This is a deterministic curse of dimensionality with respect to the number of leaves. Nevertheless, the deterministic structure avoids the simulation variance that typically affects extreme-tail estimation in Monte Carlo hierarchical aggregation methods. In practical implementations, sorting, truncation and pruning of negligible weights can reduce the computational burden while preserving the deterministic nature of the calculation.

\section{Conclusion}\label{sec:conclusion}

This paper develops an analytical framework for hierarchical copula
aggregation based on explicit finite-sum representations of aggregate
loss distributions and tail-risk measures. The main contribution is the
deterministic evaluation of quantiles, expected tail losses and
market--credit tail contributions directly from empirical marginals and
copula density kernels, without recursive copula simulation.

The results complement simulation-based hierarchical aggregation methods
by providing tractable analytical representations for both distributional
and tail-risk quantities. The compatibility theorem clarifies the
nontrivial consistency constraints linking local hierarchical copulae and
global leaf-level dependence structures. The Gaussian and mirrored
Clayton cases further illustrate how different tail-dependence regimes
induce qualitatively different asymptotic behaviours under increasing
hierarchical depth.

The asymptotic analysis highlights a fundamental distinction between
Gaussian and tail-dependent aggregation structures. In the Gaussian case,
diversification effects remain asymptotically dominant, whereas mirrored
Clayton copulae propagate upper-tail dependence recursively through the
hierarchy, potentially amplifying extreme portfolio losses and expected
tail losses at high confidence levels.

Several directions for future research remain open. A rigorous
asymptotic theory for deep hierarchical copula trees, including limit
theorems for quantiles and expected tail losses under recursive tail
dependence, appears particularly promising. Other extensions include
higher-dimensional internal aggregation nodes, dynamic copula parameters
and alternative tail-dependent copula families.

From an actuarial perspective, the framework is particularly relevant
for capital allocation, solvency analysis, reinsurance structures and
enterprise-wide risk aggregation, where transparent and reproducible
tail-risk decompositions are required
\citep{Dias2025Extended,Tasche2008,SolvencyII2009}.

\appendix
\section{Proofs}\label{app:proofs}

\subsection{Proof of Theorem~\ref{thm:compatibility}}

The boundary conditions in \eqref{eq:compatibility-boundary} follow directly from the defining marginal conditions of a copula. For the integral equation, write
\[
F^{1,2}(l_1,l_2)=\mathbb P(L_1\le l_1,L_2\le l_2).
\]
Using $L_i=L_i^{\MR}+L_i^{\CR}$ and the joint density induced by the global leaf-level copula $C$, the density of $(L_1,L_2)$ induced by $C$ is obtained by differentiating the convolution representation:
\[
\begin{split}
f^{C}_{1,2}(l_1,l_2)=&\int_{\R^2}\rho_1^{\MR}(x_1)\rho_1^{\CR}(l_1-x_1)
\rho_2^{\MR}(x_2)\rho_2^{\CR}(l_2-x_2)\\
&\quad\times c(u_1^{\MR},u_1^{\CR},u_2^{\MR},u_2^{\CR})\,\dd x_1\dd x_2,
\end{split}
\]
where
\[
(u_1^{\MR},u_1^{\CR},u_2^{\MR},u_2^{\CR})=
\big(F_1^{\MR}(x_1),F_1^{\CR}(l_1-x_1),F_2^{\MR}(x_2),F_2^{\CR}(l_2-x_2)\big).
\]
On the other hand, the hierarchical specification first induces the intermediate densities
\[
\rho_i(l_i)=\int_{\R}c_i(F_i^{\MR}(x),F_i^{\CR}(l_i-x))\rho_i^{\MR}(x)\rho_i^{\CR}(l_i-x)\,\dd x,
\]
and then couples $L_1$ and $L_2$ through the top-level copula density $c_{12}$. Hence the top-level density is
\[
f^{\mathrm{hier}}_{1,2}(l_1,l_2)=c_{12}(F_1(l_1),F_2(l_2))\rho_1(l_1)\rho_2(l_2).
\]
Expanding $\rho_1(l_1)$ and $\rho_2(l_2)$ in this last expression gives the second term inside the brackets in \eqref{eq:compatibility}. Equating $f^{C}_{1,2}$ and $f^{\mathrm{hier}}_{1,2}$ yields \eqref{eq:compatibility}.

\subsection{Proof of Propositions~\ref{prop:gaussian-total} and~\ref{prop:clayton-total}}

We prove the Gaussian case; the mirrored Clayton case is identical after replacing the Gaussian density kernel by \eqref{eq:survival-clayton-density}. Start from the bivariate aggregation operator
\[
F_L(l)=\int_{\R}\int_{-\infty}^{l-y}c(F_X(x),F_Y(y))F'_X(x)F'_Y(y)\dd x\dd y.
\]
If the marginals are empirical,
\[
F'_X(x)=\frac1n\sum_{i=1}^n\delta(x-x_i),\qquad
F'_Y(y)=\frac1n\sum_{j=1}^n\delta(y-y_j),
\]
and therefore, using the defining property of the Dirac delta,
\[
F_L(l)=\frac{1}{n^2}\sum_{i,j=1}^n \ThetaH(l-x_i-y_j)c(F_X(x_i),F_Y(y_j)).
\]
Applying this calculation first to the within-portfolio market--credit aggregation and then to the top-level portfolio aggregation gives \eqref{eq:gaussian-total}. The weights are obtained by substituting the Gaussian density kernel \eqref{eq:gaussian-density-kernel} at each node, which produces the factors $W^G_{j i h k}$. The mirrored Clayton formula follows by substituting \eqref{eq:survival-clayton-density} at each node. The repeated use of the identities
\[
\int_{\R}\delta(x-a)g(x)\dd x=g(a),\qquad
\int_{-\infty}^{l}\delta(x-a)\dd x=\ThetaH(l-a)
\]
is precisely what converts the convolution integrals into finite sums.

\subsection{Proof of Proposition~\ref{prop:var-finite}}

For a finite weighted distribution, $F_L(l)$ is a nondecreasing step function with jumps of size $p_{(m)}$ at the ordered losses $\ell_{(m)}$. The lower quantile is the first point at which the cumulative mass reaches level $\alpha$, which is exactly \eqref{eq:var-finite}.

\subsection{Proofs of Propositions~\ref{prop:gaussian-tail} and~\ref{prop:clayton-tail}}

For the Gaussian case, the joint empirical distribution of $(L^{\MR},L^{\CR})$ induced by the copula has density
\[
\frac{1}{n^2\sqrt{1-\gamma^2}}\sum_{i,j=1}^n
A_{ij}^{\gamma}
\delta(l^{\MR}-l^{\MR,i})\delta(l^{\CR}-l^{\CR,j}).
\]
The market tail contribution is
\[
\frac{1}{1-\alpha}\int\!\int_{l^{\MR}+l^{\CR}>q_\alpha(L)}l^{\MR}\,\dd F^{\MR,\CR}(l^{\MR},l^{\CR}).
\]
Substituting the empirical density above and integrating by the Dirac delta gives the first formula in \eqref{eq:gaussian-tail-contrib}; the credit contribution is analogous. The mirrored Clayton proof uses the same argument with the mirrored Clayton density weights $A_{ij}^{\zeta}$.

If the aggregate distribution has an atom at the quantile level,
only the fraction $\eta_\alpha(q)$ of the quantile mass required to
complete the tail probability $1-\alpha$ is retained. The corresponding
market and credit contributions are allocated proportionally with
respect to the copula-weighted empirical probabilities inside the
quantile atom. This yields the atom-corrected formulas
\eqref{eq:general-etl-atom-market}--\eqref{eq:general-etl-atom-credit}.

\subsection{Proof of Proposition~\ref{prop:error}}

This is the standard large-sample expression for quantile and expected tail loss estimation error. See \citet{YamaiYoshiba2002}, based on the classical order-statistic and robust-statistics arguments in \citet{Huber1981} and \citet{OrdStuart1994}. In the present paper it is applied after replacing the continuous density $F'$ by the weighted Dirac density induced by the hierarchical copula finite-sum representation.

For finite weighted distributions, the substitution of $F'_L=\sum_r p_r\delta_{\ell_r}$ into the moment integrals must use the same endpoint convention as the finite-distribution definition of expected tail loss. If there are no atoms at the quantile endpoints, the Dirac substitution gives the ordinary sums in \eqref{eq:dirac-moment-sums-simple}. If atoms are present, only the fractions of the endpoint atoms required to complete the relevant tail probabilities are included. This gives the atom-corrected upper-tail moments \eqref{eq:atom-corrected-upper-moments} and the interval moments \eqref{eq:atom-corrected-interval-moments}. Substitution of these corrected moments into the continuous benchmark formula yields \eqref{eq:finite-etl-error}.

\section*{Competing interests}
The authors declare that the views expressed in this article are their own. Simone Farinelli is affiliated with Core Dynamics GmbH. Eric Dal Moro is employed by Signal Iduna Reinsurance.

\section*{Acknowledgment on AI usage}
Parts of the exposition benefited from AI-assisted tools used to improve clarity and organization. All modelling choices, derivations, results and interpretations remain the responsibility of the authors.

\section*{Data Availability Statement}

No external datasets were used in this study. The numerical examples were generated from the analytical models described in the paper. Additional computational details are available from the corresponding author upon reasonable request.

\bibliographystyle{plainnat}
\bibliography{references_TailRiskFormulae}

\end{document}